\documentstyle[prb,twocolumn,aps]{revtex}
\input psfig
\begin{document}
\draft
\title{
Resonant Tunneling and Persistent Current of a Non-interacting and Weakly
Interacting One-dimensional Electron Gas}
\author{I. V. Krive$^{(1,2)}$ and P. Sandstr{\"o}m$^{(1)}$}
\address{$^{(1)}$Department of Applied Physics, 
Chalmers University of Technology and G{\"o}teborg University, 
S-412 96 G{\"o}teborg, Sweden\\
and $^{(2)}$B. Verkin Institute for Low Temperature Physics and Engineering, 
Academy of Sciences of Ukraine, 310164 Kharkov, Ukraine} 
\maketitle
\begin{abstract}

The persistent current for a one-dimensional ring with two
tunnel barriers is considered in the limit of
weakly interacting electrons.
In addition to a small off-resonance current, there
are two kinds of resonant behavior;
(i) a current independent of the barrier transparency (true resonance)
and (ii) a current analogous to the one for a ring with only a
single barrier (``semi''-resonance). 
For a given barrier transparency
one or the other type of resonant behavior is realized depending on
a geometric factor (ratio of interbarrier distance to ring
circumference) and on the strength of the electron-electron interaction.
It is shown that a repulsive interaction favours the``semi''-resonance
behavior.
For a small barrier transparency the ``semi''-resonance peaks are
easily washed out by temperature whereas the true resonance peaks
survive. 


\end{abstract}

\section{Introduction}
The existence of persistent currents in normal mesoscopic metal rings
was predicted quite a long time ago\cite{Gunther,Kulik.BIL,Buttiker}, 
but attracted small attention until the recent years. 
The situation changed drastically after the experimental discovery 
of persistent currents in multichannel metallic rings\cite{Levy} and
in few-channel semiconductor ballistic rings.\cite{Mailly}
The currents  measured in the disordered metallic rings\cite{Levy}
were much larger than expected and the results are still not
quite theoretically understood.
As for quantum rings with a small number of scatterers\cite{Mailly}, 
a naive model of
free electrons gives a qualitatively satisfactory explanation of
the available experimental data.
This fact puts the question why the electron-electron correlations,
which should play a significant role for the dynamics of low-dimensional
electron systems (quantum dots and quantum wires) do not affect the
persistent current in quantum rings. 
From a general point of view, a persistent current is 
due to coherence
of mesoscopic many-body systems and could be a sensitive method of 
detecting non-Fermi liquid behavior of electrons in quantum devices.

In a previous paper\cite{KriveWigner} we studied 
the persistent current in a
one-dimensional (1D) ring of strongly correlated electrons.
It is known (the Kane-Fisher effect\cite{Kane.Fisher}) that
repulsive interactions strongly suppress the tunneling in 1D electron
systems. When applied to persistent currents in a quantum ring which is 
partially
cut-off by a gate-induced potential (or by an impurity potential) this result
leads to power law suppression of the zero temperature persistent 
current\cite{Gogolin,KriveWigner} and predicts an anomalous (non-monotonic)
temperature dependence of the amplitude of the Aharonov-Bohm 
oscillations.\cite{KriveWigner,Mori}
We also showed that the anomalous temperature dependence arising
already for a short-range (Luttinger-liquid) interaction becomes
even more pronounced for long-range Coulomb forces 
(Wigner crystal).\cite{KriveWigner}

The above mentioned properties are from a theoretical point of view
consequences of the singular low momentum ($q \rightarrow 0$) behavior
of interacting electrons in one dimension. For a finite size system such 
singularities appear in the form of a power law size- and temperature
dependence. 
For the case of a single impurity the relevant
scale is the ring size $L$ and one gets a $(1/L)^{1/g}$
behavior of the persistent current amplitude at low temperatures
($g \ll 1$ is the dimensionless conductance).\cite{KriveWigner,Gogolin}
For a strong electron-electron interaction different impurities 
act non-coherently.
In this case the cut-off length is diminished ($L$ is replaced by
the average impurity spacing $\overline{l} << L$) and one can expect
an enhancement of the tunneling amplitude through each impurity by
a factor $(L/ \overline{l})^{1/g} >> 1$. Nevertheless the net current
in a ring will be strongly suppressed because of the incoherent contribution
of single-impurity trajectories to the tunneling action
$S_t^{(n)}(L) \simeq nS_t^{(1)}(\overline{l})$ (here $n = L/\overline{l},
 S_t^{(1)}$ is the one-impurity tunneling action). Hence for a strong
electron-electron interaction the persistent current in a non-ideal
ring will be much smaller than that of non-interacting electrons.

The persistent current measured in real experiments\cite{Mailly}
is of the same order as the (maximum) current of non-interacting
electrons $I_{0} \sim ev_{F}/L$ ($v_{F}$ is the Fermi velocity).
This fact means that electron-electron interactions in real quantum
rings may not be so strong ($g \sim 1$).
Therefore it is interesting and important to study
Aharonov-Bohm oscillations in a weakly interacting electron gas
where, in principle, resonant tunneling is possible.
This is the object of the present paper. More concretely we want to
understand the following points: \\
(i) How do forward and backward scattering by impurities influence
the persistent current?
\\
(ii) Is there an anomalous temperature dependence of the amplitude of 
persistent currents in the weakly interacting case? \\
(iii) What is the manifestation of resonant tunneling in persistent currents?
\\
For weak interactions it is reasonable to consider electrons as current
carrying (propagating) quasiparticles. Then the logarithmic singularities
inherent to any 1D problem of interacting gapless particles, express 
themselves in a nonperturbative renormalization of the scattering 
amplitude.\cite{MatveevYue} That is why we at first 
(by the transfer matrix method) 
will study the persistent current of non-interacting electrons in a ring
with a single (section II) and double (section III) potential barrier.
The effect of a weak electron-electron interaction, following 
Ref. \onlinecite{MatveevYue}, is taken into account by a renormalization
of the barriers. In Conclusion (section IV) we compare the predictions
of our model with the experimental results of coherent resonant tunneling
through a quantum dot.

Recently B{\"u}ttiker and Stafford \cite{Buttiker.Stafford} 
considered a mesoscopic ring with
two tunnel barriers (forming a 1D "dot" embedded in the ring). 
The analytical analysis was performed within the tunnel
Hamiltonian model describing the coupling between the states localized 
in the so formed two parts of the ring. They found a resonant behavior of
the persistent current as a function of the charge accumulated in the dot.
Here we want to describe how the persistent current depends
on the separation between the barriers and to study the temperature
behavior of the resonance peaks. 
This is why we go beyond the phenomenological treatment and use the real space
Schr{\"o}dinger formulation of the problem.

\section{Single barrier}
Let us consider a one-dimensional mesoscopic ring formed in a laterally
confined two-dimensional electron gas. The ring encloses a magnetic flux, 
$\Phi$, directed perpendicular to the plane. In addition a single tunnel
barrier is introduced into the ring. The Hamiltonian for the ring of
non-interacting electrons is
\begin{equation}
{\cal H} = \frac{p^2}{2m} + V(x)
\end{equation}
where $x$ is a spatial coordinate, $0 \leq x \leq L$ (L is the circumference
of the ring). The magnetic vector potential has been gauged
away from the Hamiltonian in a standard way\cite{ByersYang} and the magnetic
flux now enters the problem through a twisted boundary condition
\begin{equation}
\Psi(x+L) = \Psi(x) \exp \left(
2 \pi i \frac{\Phi}{\Phi_0} \right) .
\label{twistedBC}
\end{equation}
Here $\Psi$ is the electron wavefunction and $\Phi_0 = hc/e$ is the
fundamental flux quantum. 

A persistent current for a ring in contact with an electron reservoir
is calculated from the expression
\begin{equation}
I = -c \frac{ \partial \Omega }{\partial \Phi} ,
\end{equation}
where $\Omega$ is the thermodynamic potential. In order to 
obtain the energy spectrum for the electron system, we have used
the transfer matrix method. The scattering of electrons by a barrier
is described by a transfer matrix
\begin{equation}
\hat{T} = \left(
\begin{array}{cc}
1/t & r/t \\
r^{*}/t^{*} & 1/t^*
\end{array}
\right), \quad |r|^2 + |t|^2 =1 ,
\label{transfer}
\end{equation}
where t (r) is the forward (backward) scattering amplitude which was
parameterized as 
$t(p) = \sqrt{T_{t}(p)} e^{i\delta_{+}(p)}$.
Here $T_t(p)$ is the transmission probability and $\delta_{+}(p)$ the 
forward scattering phase. The wavefunction at the left side of the barrier
can be written as $\Psi(x) = Ae^{ipx} + Be^{-ipx}$ and at the right side
as $\Psi(x) = Ce^{ipx} + De^{-ipx}$.
The matrix equation 
\begin{equation}
\left(
\begin{array}{c}
C \\ D
\end{array}
\right)
=
\hat{T}
\left(
\begin{array}{c}
A \\ B
\end{array}
\right)
\end{equation}
together with Eq.~(\ref{twistedBC}) lead to the equation for allowed
momenta\cite{Gogolin}
\begin{equation}
\cos(pL+\delta_{+}(p)) = \sqrt{T_{t}(p)} \cos\left(2 \pi \frac{\Phi}{\Phi_0} \right).
\label{singlespec}
\end{equation}
Since a persistent current is sensitive only to the energy spectrum
at or close to the Fermi energy, we have made the following assumptions
\begin{equation}
\begin{array}{ll}
T_{t}(p) \Rightarrow T_{t} = const \\
\delta_{+} (p) \Rightarrow \delta_{+} sign (p)
\end{array}
\label{assumption}
\end{equation}
where $T_t$ and $\delta_+$ are the values of the transmission probability
and the forward scattering phase at the Fermi surface. The potential
is assumed to be smooth at the Fermi level. 

The persistent current can now be calculated analytically for arbitrary
values of temperature and the barrier transparency.
The result is
\begin{eqnarray}
&& I(\Phi,T,T_t) = 4 \pi I_{0} \frac{T}{\Delta} 
\frac{ \sin (2 \pi \Phi/\Phi_0) }{ \sqrt{T_t^{-1} - \cos^{2}(2 \pi \Phi/\Phi_0)} } \times \label{currentone} \\
&& \sum_{k=1}^{\infty} \frac{\sin \left\{k \arccos \left[\sqrt{T_t}\cos(2 \pi \Phi/\Phi_0) \right] \right\} \cos \left\{ k(\delta_{+} + 2 \pi \mu/ \Delta ) \right\} }{\sinh(k 2 \pi^{2} T/\Delta)} , \nonumber
\end{eqnarray}
where $I_{0} \equiv ev_{F}/L$, $\Delta \equiv 2 \pi \hbar v_{F}/L$ is the 
level spacing at the Fermi energy, $v_{F}$ is the Fermi velocity and
$\mu$ is the chemical potential of the 2D electron gas.
As seen from Eq.~(\ref{currentone}), the only effect of
the forward scattering phase, $\delta_{+}$, is to shift the chemical
potential $\mu \rightarrow \mu_{eff} = \mu + \Delta \delta_{+} /2 \pi$.

When the phase of a charged particle's wave function is shifted
due to the presence of an electrostatic potential,
Aharonov-Bohm like oscillations arise.
This 
is sometimes called the electrostatic Aharonov-Bohm effect
(see e.g. Ref. \onlinecite{Washburn}) and can in our approach
be modelled by a scattering with perfect transmission $T_{t} = 1$.
Contrary to the interaction with a vector potential, the phase shift
induced by purely forward scattering is identical both for the 
clockwise and the counterclockwise moving particles
(see Eq.~(\ref{singlespec})). It results in a homogeneous shift
of energy levels of electrons in a ring. Each time when a level crosses
the Fermi surface of the reservoir, an extra electron enters (or leaves)
the ring and the grand potential $\Omega$ oscillates. It is evident
that these oscillations are completely identical to the ones caused by
varying the chemical potential, $\mu$.

As follows from Eq.~(\ref{currentone}) the crossover temperature 
$T^{*}=\Delta/2 \pi^{2}$ (the temperature is suppressed at temperatures
$T \agt T^{*}$) does not depend on the barrier characteristics.
It coincides with the crossover temperature for 
free electrons in a perfect (impurity free) ring in contact with an
electron reservoir 
(see e.g. Ref. \onlinecite{Zvyagin}). At low temperatures $T \ll T^{*}$
Eq.~(\ref{currentone}) simplifies to
\begin{eqnarray}
I(\Phi,T \rightarrow 0) = -\frac{1}{\pi} I_{0} 
\frac{\sin(2 \pi\Phi/\Phi_0)}{\sqrt{T_t^{-1} - \cos^{2}(2 \pi \Phi/\Phi_0)}}
\times \nonumber \\
\left\{
\begin{array}{rr}
\arccos \left[\sqrt{T_t}\cos(2 \pi \Phi/\Phi_0) \right] \quad N\; odd \\
- \pi + \arccos \left[\sqrt{T_t}\cos(2 \pi \Phi/\Phi_0) \right] \quad N\; even 
\end{array}
\right.
\end{eqnarray}
which is the zero temperature result given in 
Ref.~\onlinecite{Gogolin}
(here $N$ is the total number of conduction electrons in the ring).
For a small transparency of the barrier, the current is small 
$\propto \sqrt{T_{t}}$ and takes a sinusoidal form
\begin{equation}
I(\Phi) \propto I_{0} \sqrt{T_{t}} \sin \left( 2 \pi \frac{\Phi}{\Phi_0} \right) \quad T \ll T^{*} .
\label{harmonic}
\end{equation}
The same conclusion is true also for high temperatures
$T \agt T^{*}$ when additional exponential damping appears
\begin{equation}
I(\Phi) = \frac{4}{\pi} I_{0} \sqrt{T_{t}} \frac{T}{T^{*}}
e^{-T/T^{*}} \sin \left(2 \pi \frac{\Phi}{\Phi_0}\right) \cos \left(2 \pi \frac{\mu_{eff}}{\Delta} \right) .
\end{equation}

Let us now study the persistent current of 1D weakly interacting electrons
in the presence of a single barrier. 
For weakly interacting electrons the effect of interactions on scattering can 
be taken into account by a barrier renormalization.
The incident electron is scattered not only by the bare external but also
by the potential induced by charge density oscillations --- Friedel
oscillations --- caused by backward scattering.\cite{Kane.Fisher}
For perfect transmission, $T_{t} = 1$, the Friedel oscillations are absent 
and electron-electron interactions do not 
affect electrostatic Aharonov-Bohm oscillations.
For the case of a non-perfect ring, the transmission probability is strongly 
renormalized.\cite{Kane.Fisher,MatveevYue} Using the results of 
Ref. \onlinecite{MatveevYue} we get the renormalized transmission probability,
$T_{t}^{R}$, for a system of finite size $L$
(energy level spacing $\Delta \sim \hbar v_{F}/L$) and at temperature $T$
as follows

\begin{equation}
T_t^{R} = 
\left\{
\begin{array}{ll}
T_{t} \left( \frac{\Delta}{D} \right) ^{2 \gamma} / 
\left\{(1-T_{t}) + T_{t} \left( \frac{\Delta}{D} \right) ^{2 \gamma} \right\},
\quad T \ll \Delta \\
T_{t} \left( \frac{T}{D} \right) ^{2 \gamma} /
\left\{(1-T_{t}) + T_{t} \left( \frac{T}{D} \right) ^{2 \gamma} \right\},
\quad T \agt \Delta .
\end{array}
\right.
\label{renormbarrier}
\end{equation}
In Eq.~(\ref{renormbarrier}) $D \sim \epsilon_{F}$ is a bandwidth 
cutoff and $\gamma$ characterizes
the strength of the electron-electron interaction 
($\gamma = V_{0}/2 \pi \hbar v_{F}$ if only forward scattering is 
included\cite{MatveevYue}, here
$V_{0}$ is the forward scattering amplitude of the electron-electron
interaction). 
The interaction parameter $\gamma$ can be expressed in terms of
the Luttinger liquid stiffness constant 
$\alpha = v_{F}/s$ ($s$ is the plasmon velocity) according to
\begin{equation}
\gamma = \frac{1}{2} \left( \frac{1}{\alpha^{2}} - 1 \right) .
\end{equation}
For weakly interacting electrons ($\alpha \rightarrow 1$) this relation
can simply be written as $\gamma \simeq \alpha^{-1} -1$. 
The interesting limiting case of Eqs.~(\ref{currentone}), (\ref{renormbarrier})
is the one of small barrier transparency ($T_{t} \ll 1$).
In this limit and at low temperatures ($T \ll T^{*}$) the persistent
current takes the form (see also \cite{Gogolin})
\begin{equation}
I_{int} (\Phi) \simeq \pm \frac{1}{2} I_{0} \sqrt{T_{t}}
\left( \frac{\Delta}{\epsilon_{F}} \right)^{\frac{1}{\alpha}-1}
\sin \left( 2 \pi \frac{\Phi}{\Phi_{0}} \right) .
\label{currentint}
\end{equation}
The current is positive (paramagnetic response) for 
$0 < \mu_{eff} < \Delta/4$ and negative (diamagnetic response)
for $\Delta/4 < \mu_{eff} < \Delta/2$.
We note that at low temperatures the oscillations of the persistent current
as a function of $\mu$ are rectangular 
(the parity effect, see Refs. \onlinecite{Loss.prl,Kusmartsev})
and for a small barrier transparency they have equal steps of para-
and diamagnetic responses
$\Delta \mu^{eff}_{p} = \Delta \mu^{eff}_{d} = \Delta/4$.
The interaction dependent factor $(\Delta/\epsilon_{F})^{1/\alpha - 1}$
in Eq.~(\ref{currentint}) looks similar to the strongly interacting case,
$\alpha \ll 1$.\cite{KriveWigner}
However now $\alpha \alt 1$ (more precisely $1 - \alpha \ll 1$)
and Kane-Fisher's suppression is not pronounced.
For high temperatures the current is
\begin{eqnarray}
I_{int} (\Phi) \sim && I_{0}
\sqrt{T_{t}} \left( \frac{\Delta}{\epsilon_{F}} \right)^{\frac{1}{\alpha}-1}
\left( \frac{T}{\Delta} \right)^{\frac{1}{\alpha}} \times \nonumber \\
&&
e^{-T/T^{*}} \sin \left(2 \pi \frac{\Phi}{\Phi_{0}} \right)
\cos \left(2 \pi \frac{\mu_{eff}}{\Delta} \right) .
\end{eqnarray}
Here again we have reproduced the power law temperature factor
$(T/\Delta)^{1/\alpha} \gg 1$.\cite{Kane.Glazman,KriveWigner}

It is interesting to note that 
Eqs.~(\ref{currentone}), (\ref{renormbarrier}) do not predict an
anomalous temperature dependence for the persistent current
of weakly interacting electrons (it is beyond the accuracy which
Eq.~(\ref{renormbarrier}) holds with), in contrast to the case of 
strongly repulsive electrons where a sharp maximum develops with 
temperature.\cite{KriveWigner}

\section{Double barriers}
\begin{figure}
\centerline{\psfig{figure=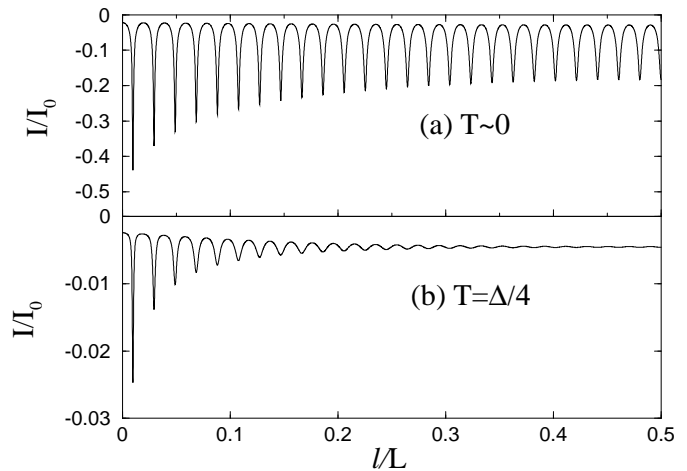,width=9.5cm}}
\vspace*{2mm}
\caption{ \protect{\label{fig:highlowT}}
Persistent current at low (a) and high (b) temperatures
as a function of the 
dimensionless distance $l/L$
between the two barriers in the ring ($0 \leq l \leq L/2$). 
The transparency of each barrier is $T_{t} = 0.1 \ll 1$,
the flux penetrating the ring is $\Phi = \Phi_{0}/4$ and the total
number of (spinless) electrons is $N=51$ (odd).
The peaks are equally spaced, $\Delta l = \lambda_{F}/2$.
When the distance between the barriers are small, $l/L \ll 1$, the peaks
correspond almost to true resonances 
(the current does not dependent on $T_{t}$).
These peaks survive at finite ($T \sim \Delta$) temperatures.
When the two barriers are far apart from each other, $l/L \sim 1/2$,
the amplitudes of the peaks at low temperatures are only 
an order $1/ \protect\sqrt{T_{t}}$ larger than
the off-resonant current.
These peaks are easily washed out by temperature
(see Eq.~(\protect\ref{crossoverT})).
}
\end{figure}

When two barriers are introduced into the ring, the possibility of
resonant behavior opens up. We will here assume that the two barriers
are identical, each described by the same transfer matrix, $\hat{T}$
(see Eq.~(\ref{transfer})). 
The electron wavefunctions now have to be matched at two points in the
ring. Together with twisted boundary conditions, one gets 
after straightforward calculations the equation for allowed momenta
(non-interacting electrons)
\begin{eqnarray}
T_{t}(p) \cos \left(2 \pi \frac{\Phi}{\Phi_{0}} \right) = &&
(1 - T_{t}(p)) \cos \left[ p L \left( 1 - 2 l/L \right) \right] + \nonumber \\
&& \cos \left( pL + 2 \delta_{+}(p) \right)
\label{spectrumtwo}
\end{eqnarray}
where $l$ is the distance between the tunnel barriers ($0 \leq l \leq L/2$).
Again we let the
transmission probability $T_{t}$ and the forward scattering phase 
$\delta_{+}$ take their respective values at the Fermi surface
(see Eq.~(\ref{assumption})). 
Generally the persistent current can be expressed in the following form
\begin{equation}
I=-I_{0} \sum_{ \{p_{n}\} } f_{FD}(p_{n}) \frac{1}{\hbar v_{F}} 
\left.
\frac{\partial \epsilon(p)}{\partial p}
\right| _{p=p_{n}}
v(p_n)
\label{generalI}
\end{equation}
where $f_{FD}$ is the Fermi-Dirac distribution function,
$\epsilon (p)$ is the dispersion relation ($\epsilon (p) = p^{2}/2m$) and 
$v \equiv \partial (pL) / \partial (2\pi\Phi/\Phi_{0})$
is the dimensionless ``band velocity'' which for the spectrum
Eq.~(\ref{spectrumtwo}) looks like
\begin{equation}
v = \frac{T_{t} \sin(2 \pi \Phi/\Phi_{0})}
           {(1-T_{t}) \sigma \sin(2 \pi x \sigma) + \sin(2\pi x + 2\delta_{+})}
\label{velocity}
\end{equation}
where $\sigma \equiv 1 - 2l/L$ and $x \equiv pL/2 \pi$.
By solving Eq.~(\ref{spectrumtwo}) numerically one
can calculate the persistent current, Eq.~(\ref{generalI}), for the 
whole range of parameters entering Eq.~(\ref{spectrumtwo}).
For low transparencies, $T_{t} \ll 1$, and at low temperature
($T \rightarrow 0$), the dependence of the persistent current amplitude
on the dimensionless interbarrier distance $l/L$ is shown in 
Fig.~\ref{fig:highlowT}a
(for a magnetic flux $\Phi/\Phi_{0} = 1/4$). The plot demonstrates
a typical resonance behavior with sharp resonance peaks and small
off-resonance current $I_{off} \propto (\sqrt{T_{t}})^{2} \sim T_{t} \ll 1$.
The distinctive feature of the depicted resonant oscillations as compared
to the well-known behavior of a double barrier introduced into an open system,
is the smooth modulation of the resonance peaks. They
are almost true resonances at small distances $l/L \ll 1$
(for $\Phi/\Phi_{0} = 1/4$ the persistent current for an odd number of 
electrons
in a perfect ring should be $I/I_{0} = - 2 \Phi/\Phi_{0} = -0.5$).
With the increase of the interbarrier distance $l$, the amplitude of
the peaks decreases and saturates at the value $I/I_{0} \sim \sqrt{T_{t}}$
(in what follows we will call these peaks as semi-resonance peaks).
\footnote{In the tunnel Hamiltonian model, semi-resonance oscillations
of zero temperature persistent current were predicted in 
Ref.~\onlinecite{Buttiker.Stafford}.
An analogous effect leading to giant Josephson current in a 
superconducting tunnel junction was studied in Ref.~\onlinecite{Wendin}.}

At finite temperatures the resonance oscillations survive for $l/L \ll 1$
whereas the semi-resonance peaks vanish (see Fig.~\ref{fig:highlowT}b). 
Such an unusual
manifestation of resonant tunneling in a closed system (a ring)
demands a physical explanation. Therefore we analyze Eq.~(\ref{spectrumtwo})
analytically for different ranges of parameters to describe different
regimes of persistent current oscillations in a ring with double barriers.


It is evident from Eq.~(\ref{spectrumtwo}) that for special 
values of barrier separation, $l$,
there are momenta, ${p_{n}}$, which do not depend on the barrier
transparency. This resonance spectrum can be found from a particular 
solution of 
Eq.~(\ref{spectrumtwo})
\begin{equation}
\left\{
\begin{array}{cc}
\cos(2 \pi x \sigma) + \cos(2 \pi x + 2 \delta_{+}sign(x)) = 0 \\
\cos(2 \pi \Phi/\Phi_0) + \cos(2 \pi x \sigma) = 0 \quad .
\end{array}
\right. 
\label{resonancespec}
\end{equation}
The spectrum of true resonance
which follows from Eq.~(\ref{resonancespec}) 
is the same as for a ring without scatterers
\begin{equation}
x_{n} = \frac{\Phi}{\Phi_{0}} + n - \frac{1}{\pi} \delta_{+} sign(n)
\quad n=0, \pm 1,...
\label{xn}
\end{equation}
By substituting Eq.~
(\ref{resonancespec}) into Eq.~(\ref{velocity})
one gets the dimensionless derivatives of the energy levels over flux
(``velocities'') as follows
\begin{equation}
v^{(\pm)} = \frac{T_{t}}{1 \pm \sigma (1 - T_{t})} .
\label{velocity.pm}
\end{equation}
The two signs in Eq.~(\ref{velocity.pm}) correspond to two different
sets of roots of Eq.~(\ref{resonancespec}) for the ``lengths''
$\sigma_{k}$

\begin{equation}
x_{n} \left( 1 \pm \sigma_{k}^{(\mp)} \right) = k - 
\frac{\delta_{+}}{\pi} sign(k) \quad k = 0, \pm 1, ...
\label{roots}
\end{equation}
Let us now consider the case of a small barrier transparency
($T_{t} \ll 1$). From Eq.~(\ref{harmonic}) it is evident that off resonance, 
the persistent current should be suppressed by the order 
$\left( \sqrt{T_{t}} \right)^{2} = T_{t}$.
The velocity $v^{(+)}$ corresponds to such a small persistent current
($\sim T_{t}$) and is thus of no interest for us.
On the contrary, the velocity $v^{(-)}$ in the limit $\sigma \rightarrow 1$
(i.e. $l \rightarrow 0$) does not depend on $T_{t}$ ($v^{(-)} \rightarrow 1$).
It means that for particular (resonant) lengths $\sigma_{k}^{(-)}$ 
satisfying Eq.~(\ref{roots}) and the condition $1 - \sigma_{k}^{(-)} \ll 1$,
we will have resonant persistent current in a double barrier ring.

Since the spectrum for resonant tunneling, Eq.~(\ref{xn}), is the same as 
for a perfect ring, the temperature dependence of resonant persistent current 
(as well as the dependence of current on the forward scattering phase
$\delta_{t} = 2 \delta_{+}$) is described by a standard formula
\begin{equation}
I_{r} (\Phi, T, \delta_{t}) = 4 \pi I_{0} \frac{T}{\Delta}
\sum^{\infty}_{k=1} 
\frac{
\sin \left(k 2 \pi \frac{\Phi}{\Phi_{0}} \right)
\cos \left\{k \left(\delta_{t} + 2 \pi \mu/\Delta \right) \right\}}
{\sinh \left(k 2 \pi^{2} T/ \Delta \right) }
\end{equation}
so that the crossover temperature is $T^{*} = \Delta / 2 \pi^{2}$.
We have already seen from Eq.~(\ref{currentone}), that for a single barrier
(off resonance situation) the crossover temperature also coincides
with $T^{*}$. It means that though with the increase of temperature, the 
persistent current amplitude is decreased, the ratio of on- to off
resonance current does not depend on temperature. This fact explains
why the resonant oscillations survive even in the high temperature
region.

What is the distance between the resonance peaks?
It is evident that the resonant structure, being a consequence of
quantum mechanical interference, should be periodic in $\lambda/2$,
where $\lambda=h/p$ is the de Broglie wavelength. For persistent current 
the relevant energy scale is the Fermi energy. Therefore one can expect
the resonance peak spacing to be equal to $\lambda_{F}/2$ 
($\lambda_{F} = 2 \pi \hbar/p_{F}$ is the Fermi wavelength).
It is easy to confirm this conclusion by a direct calculation using
Eq.~(\ref{roots}) for resonant tunneling (upper sign).
By noticing that the main contribution to the persistent current
comes from the energy levels at the Fermi ``surface'' we get for
the peak positions the simple expression
\begin{equation}
l_{n}^{(r)} = \frac{\lambda_{F}}{2} 
\left( \frac{1}{2} + n \right) ,  \quad n = 0, 1, 2, ...
\label{spacing}
\end{equation}
For mesoscopic rings ($N \gg 1$) the positions of the resonance peaks
neither depend on flux, $\Phi$, nor on the transmission probability, $T_{t}$.
The peak spacing is inversely proportional to the electron density
$\Delta l = 1 / \rho$.

Now we proceed to the case when the two barriers are far apart from
each other ($l \sim L/2$). When $l/L$ is not small, both solutions of
Eq.~(\ref{resonancespec}) give a small (off resonance) current 
($\propto T_{t}$). To explain the relatively large $\sqrt{T_{t}}$-peaks
of Fig.~\ref{fig:highlowT}a one should look for another solution 
to the equation for
the spectrum. As follows from Eq.~(\ref{spectrumtwo}),
the off-resonance situation takes place when the cosine in the first
term of the righthand side of the equation is small. In the opposite limit
\begin{equation}
\cos \left(2 \pi x \sigma \right) = \pm 1
\label{short}
\end{equation}
the spectrum $x^{(s)}$ reads
\begin{equation}
\cos \left(2 \pi x^{(s)} + 2 \delta_{+} \right) =
\mp 1 + T_{t} \left\{ 
1 + \cos \left( 2 \pi \frac{\Phi}{\Phi_{0}} \right)
\right\} .
\label{semispectrum}
\end{equation}
The ``velocity'' for the semi-resonance peaks has the following form
\begin{eqnarray}
v^{(s)} = && \frac{T_{t} \sin(2 \pi \Phi/\Phi_{0})}
{\sqrt{1 - \left[\pm 1 - T_{t}(1 + \cos(2 \pi \Phi/\Phi_{0})) \right]^{2}}} 
\stackrel{T_{t} \ll 1}{\simeq} \nonumber \\
&& \sqrt{T_{t}} \frac{\sin(2 \pi \Phi/\Phi_{0})}{\sqrt{2(1 \pm \cos(2\pi \Phi/\Phi_{0}))}}
\label{vel}
\end{eqnarray}
Note here the square-root dependence of the velocity on the
transparency $T_t$ implying the same dependence for the 
persistent current (see also Ref.~\onlinecite{Buttiker.Stafford}).

Using the spectrum, Eq.~(\ref{semispectrum}), it is easy to calculate the 
persistent current at finite temperatures. It can be shown that the
``$\pm$'' signs in Eq.~(\ref{semispectrum}) correspond to a ring with
an odd (``$+$'') or even (``$-$'') number of spinless electrons.
The parity effect can as usual be reproduced by a shift of the magnetic
flux $\Phi/\Phi_{0} \Rightarrow \Phi/\Phi_{0} + 1/2$ in the final
expressions (see e.g. review \cite{Zvyagin}). 
So in what follows we will for definiteness consider an odd number of
electrons in the ring.

We omit the straightforward calculations of the persistent current,
using the spectrum Eq.~(\ref{semispectrum}) and the Fermi-Dirac
distribution function. For small barrier transparencies ($T_{t} \ll 1$)
the final expressions for the persistent current of semi-resonance peaks
takes the form
\begin{eqnarray}
I^{(s)}(\Phi)= && - \frac{2}{\pi} I_{0} \sqrt{T_{t}}
\frac{ \sin \left(2 \pi \Phi/\Phi_{0} \right)}
     {\sqrt{ 2 \left(1 + \cos \left(2 \pi \Phi/\Phi_{0} \right) \right) }}
\frac{T}{T^{*}}
\times
\nonumber \\
&& 
\sum^{\infty}_{k=1} 
\frac{ \sin \left\{k \sqrt{T_{t} 2 \left(1+\cos \left(2 \pi \Phi/\Phi_{0} \right) \right) } \right\} }
     {\sinh \left(k T/T^{*} \right) } .
\label{semicurrent}
\end{eqnarray}
At small temperatures $T \ll T^{*}$ the summation in Eq.~(\ref{semicurrent})
can be replaced by integration and the desired current is

\begin{eqnarray}
I^{(s)}(\Phi)= && -I_{0} \frac{\sqrt{T_{t}} \sin(2 \pi \Phi/\Phi_{0})}
                          {\sqrt{2(1+\cos(2 \pi \Phi/\Phi_{0}))}} \times
\nonumber \\
&& \tanh \left\{ \frac{\pi}{2}
\frac{T^{*} \sqrt{2 T_{t}(1+\cos(2 \pi \Phi/\Phi_{0}))}}{T} \right\} .
\label{lowTsemicurrent}
\end{eqnarray}
We see that the crossover temperature $T_{s}^{*}$ for semi-resonant
currents depends on the transparency and for small $T_{t}$ it is much
smaller than $T^{*}$
\begin{equation}
T_{s}^{*} = \frac{\pi}{2} T^{*} 
\sqrt{2 T_{t} (1 + \cos(2 \pi \Phi/\Phi_{0}))} 
\stackrel{\Phi/\Phi_{0} \neq 1/2}{\sim} \sqrt{T_{t}} T^{*} \ll T^{*} .
\label{crossoverT}
\end{equation} 
The zero temperature ($T \ll T_{s}^{*}$) current is
($N$ is odd and the current is diamagnetic)
\begin{equation}
I^{(s)}(\Phi) \stackrel{T \rightarrow 0}{\sim} - I_{0} \sqrt{T_{t}}
\sin \left( \pi \frac{\Phi}{\Phi_{0}} \right)
sign \left\{ \cos \left( \pi \frac{\Phi}{\Phi_{0}} \right) \right\} .
\end{equation}
For an even number of electrons in the ring, the magnetic flux should be
shifted by $\Phi_{0}/2$ and the response becomes paramagnetic as dictated
by the parity effect for spinless fermions.


At temperatures $T \gg T_{s}^{*}$ the current in Eq.~(\ref{lowTsemicurrent}),
$I^{(s)} \propto I_{0} T_{t} T^{*}/T$, is of the order of the non-resonant
current and the high temperature semi-resonance oscillations disappear
(see Fig.~\ref{fig:highlowT}b).
 
It is worthy to note here that analogous estimations hold for the 
temperature dependence of paramagnetic current in a perfect ring at low 
fluxes ($\Phi/\Phi_{0} \ll 1$)
\begin{eqnarray}
I(\Phi) = && \frac{2}{\pi} I_{0} \frac{T}{T^{*}}
\sum^{\infty}_{k=1} \frac{ \sin \left(2 \pi k \Phi/\Phi_{0} \right)}
                         {\sinh \left( k T/T^{*} \right) }
\stackrel{\Phi/\Phi_{0} \ll 1, T \ll T^{*}}{\sim}
\nonumber \\
&& I_{0} \tanh \left( \frac{T^{*}}{T} \pi^{2} \frac{\Phi}{\Phi_{0}} \right) .
\end{eqnarray}
This analytic formula explains how the jump of current
$\Delta I = 2 I_{0}$ at $T \rightarrow 0$ and $\Phi = 0$ is smeared
by temperature.

The position of the semi-resonance peaks (Fig.~\ref{fig:highlowT}a) 
can be found from
Eqs.~(\ref{short}), (\ref{semispectrum})
\begin{equation}
l_{n}^{(s)} = \frac{L}{2} - n \frac{\lambda_{F}}{2} - 
\frac{\lambda_{F}}{8} \left[1+(-1)^{N} \right] , \quad n=0, 1, 2, ...
\label{semipos}
\end{equation}
The spacing of the semi-resonance peaks $\Delta l = \lambda_{F}/2$ 
is the same as for the resonance peaks, but unlike the case of 
``true'' resonances their positions depend on the parity of the
total number of electrons in the ring (see the last term in 
Eq.~(\ref{semipos}). For mesoscopic rings ($N \gg 1$) with not a too
small transparency, the resonance peaks smoothly transform to the
semi-resonance ones (Fig.~\ref{fig:highlowT}a).

Now we can estimate the crossover length, $l_{c}$, from resonance to
semi-resonance behavior. By equating the two ``velocities'' of 
Eqs.~(\ref{velocity.pm}) and (\ref{vel}) one gets
\begin{equation}
l_{c}/L \sim \sqrt{T_{t}} \ll 1
\end{equation}
For small barrier transparencies, the range of ``true'' resonance oscillations
(see Fig.~\ref{fig:highlowT}a) is very narrow and can be neglected. 
One can say that a
strong double barrier in a closed system leads only to semi-resonance peaks
($I_{peak} \approx I_{0} \sqrt{T_{t}}$ at T=0) which easily are washed out
by temperature. It means that at not too small temperatures and for
strong enough barriers, any resonant behavior of persistent current is 
improbable. The inclusion of interactions only strengthens this general
conclusion (see below).

It is useful to present a simple qualitative explanation of the 
appearance of the semi-resonance peaks. Let us consider a ring divided
by two strong barriers into two independent segments.
When $l=L/2$ (the barriers are situated diametrically opposite)
the segments are identical and in the absence of tunneling 
($T_{t} \rightarrow 0$) all energy levels are doubly degenerated.
Tunneling lifts this degeneracy and gives rise to an energy splitting
which is proportional to $\sqrt{T_{t}}$. This very energy splitting
also determines the width of the band structure of energy levels 
as a function of flux $\Phi$. The persistent current, being
proportional to the band velocity at the Fermi level, is also of
the order of $\sqrt{T_{t}}$. So in order to have large 
(proportional to $\sqrt{T_{t}}$ rather than $T_{t}$)
persistent current, one needs to arrange a
situation where in the absence of tunneling there is a degeneracy
at the Fermi energy.
If $l \neq d = L/2 -l$ and $\Delta l = |l - d| \ll l,d$, it is easy to show
starting from a spectrum in absence of tunneling
$p_{k}^{l(d)} = \frac{\pi}{l(d)}(k+1/2)$ ($k$ is an integer) that the
resonance condition for mesoscopic rings ($N \gg 1$) is satisfied for the set
of $\Delta l_{n} = n/\rho_{0}$ ($\rho_{0}$ is the one-dimensional electron
density, $n = 1, 2, 3, ...$). We again come to the peak spacing 
$\Delta l = \lambda_{F}/2$.

\begin{figure}
\centerline{\psfig{figure=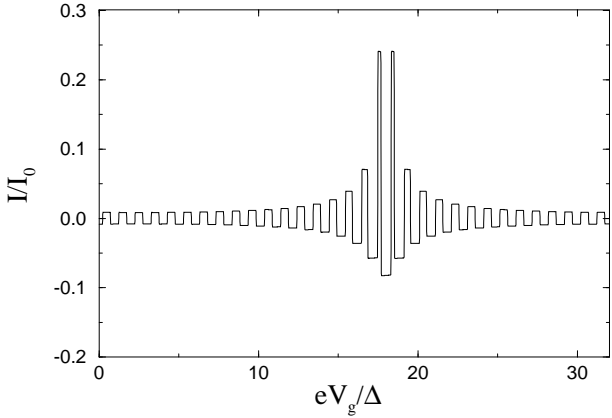,width=9cm}}
\vspace*{2mm}
\caption{ \protect{\label{fig:rectangular}}
Low-temperature persistent current as a function of normalized gate voltage.
$\Delta$ is the energy level spacing. 
The transparency of each barrier is $T_{t} = 0.1 \ll 1$ and
the flux penetrating the ring is $\Phi = \Phi_{0}/20$.
The number of electrons changes with the gate voltage and the current
exhibit dia$\leftrightarrow$paramagnetic oscillations.
For certain gate voltages the density of electrons for a fixed
ratio $l/L$ satisfies the resonance condition and the persistent current
peaks.
}
\end{figure}

Since the spacing in question depends on the electron density, one can 
expect resonant-like behavior of the persistent current amplitude as
a function of the gate voltage. To consider (qualitatively) this
effect we start (in the general case) from a non-resonant situation
($l, \rho_{0}$) corresponding to a small current ($I \sim T_{t} \ll 1$).
By varying the gate voltage, one can change the number of electrons
on the ring. Hence the persistent current, still remaining small,
should exhibit dia$\leftrightarrow$paramagnetic oscillations which 
are rectangular at low temperatures\cite{unpublished}.
If $T_{t} \ll 1$ the plateau for diamagnetic response and the plateau for
paramagnetic response are practically equal (see Eq.~(\ref{currentint})).
With a further increase of the number of 
electrons on the ring, we come at last to a
density $\rho$ which matches at fixed $L$ and $l$ the resonance condition.
As a result the persistent current peaks. The numerical calculations           confirm this scenario (see Fig.~\ref{fig:rectangular}).

\begin{figure}
\centerline{\psfig{figure=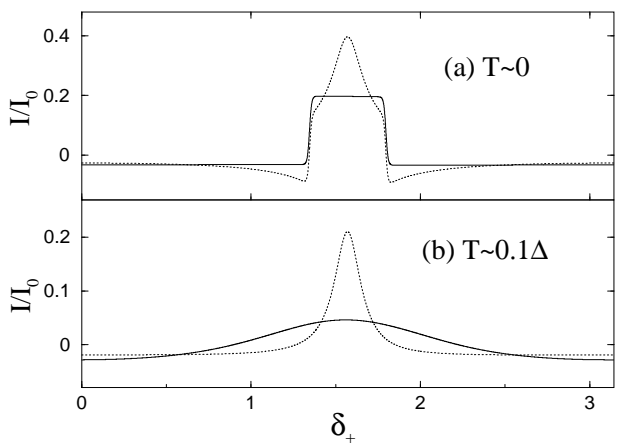,width=9.5cm}}
\vspace*{2mm}
\caption{ \protect{\label{fig:deltaplus}}
Persistent current at low (a) and finite (b) temperatures as a
function of the forward scattering phase which models a plunger 
gate voltage (see text). 
The transparency of each barrier is $T_{t} = 0.1 \ll 1$,
the flux penetrating the ring is $\Phi = \Phi_{0}/4$ and the total
number of electrons is $N=51$.
Two different interbarrier distances ($l/L=1/N$ and $l/L=24/N$)
are considered.
The larger (smaller) distance corresponds to the semi- (true-) 
resonant region.
At low temperatures semi-resonant behavior (solid line) is characterized 
by a broad step-like burst of the current whereas the true-resonance peaks
(dotted line) still remain sharp. Temperature strongly suppresses
and widens the semi-resonance bursts of current while the
true-resonance peaks, though suppressed in amplitude,
keep their shape.
}
\end{figure}

In our approach we can model the influence of a plunger gate voltage on the
persistent current by varying the forward scattering phase, $\delta_{+}$.
One could expect that starting from an off-resonant current at
$\delta_{+} = 0$, one obtains a resonant current for certain
values of $\delta_{+}$. This behavior is depicted in 
Fig.~\ref{fig:deltaplus} for two different cases 
(with barrier separations corresponding to true- and semi-resonant regions).
Now semi-resonant behavior is characterized by a broad step-like
burst of the current (at $T=0$) whereas the true-resonance peaks
still remain sharp (see Fig.~\ref{fig:deltaplus}a). 
Temperature strongly suppresses
and widens the semi-resonance bursts of current while the
true-resonance peaks, though suppressed in amplitude,
keep their shape (see Fig.~\ref{fig:deltaplus}b).

Until now we have studied the behavior of noninteracting electrons in 
a double barrier ring. For an open system it is known\cite{Kane.Furusaki}
that resonant tunneling takes place even for interacting electrons if the
interaction is not strong. It is evident in the approach developed in 
Ref. \onlinecite{MatveevYue} that weak interactions can not affect resonant
tunneling (for $T_{t}(p_{r}$)=1 there are no Friedel oscillations causing
a barrier renormalization).

Semi-resonance peaks ($\propto \sqrt{T_{t}}$) appear in the situation 
when the barriers are situated far from each other 
($l \sim L/2 \gg \lambda_{F}$).
In this case the effect of electron-electron interactions can be
accounted for by an independent renormalization of the transparency 
of each individual barrier, Eq.~(\ref{renormbarrier}). 
According to this formula, both the
semi-resonance peaks and the non-resonant current are suppressed by 
interactions. At $T=0$ their ratio 
$I^{(s)}/I_{off} \sim \left(T_{t}^{R} \right)^{-1/2} \sim 
\frac{1}{\sqrt{T_{t}}} 
\left( \frac{\epsilon_{F}}{\Delta} \right)^{1/\alpha - 1}$
will increase with the increase of interactions ($\alpha=1$ for the
noninteracting case and $\alpha < 1$ for repulsive interactions).
The peaks become more pronounced and the range of semi-resonance oscillations
is extended (the renormalized crossover length
$l_{c}^{R} \sim l_{c} \left( \frac{\Delta}{\epsilon_{F}} \right)^{1/\alpha-1}
\ll l_{c}$).
However the repulsive interaction also diminishes the crossover temperature
$T_{s}^{*}$, Eq.~(\ref{crossoverT}), of semi-resonance oscillations.
Therefore at finite temperatures the influence of (even weak) interactions
gives a tendency to prevent the appearance of all kinds of resonant effects.
For strong interactions different barriers act non-coherently already at
zero temperature. Nevertheless unlike the case of strong interaction,
weak (or moderate) electron-electron repulsion allows in principle the
current of a ring with impurities to be resonant.

\section{Conclusions}
Recently coherent resonant tunneling through a quantum dot
was observed in several sets of experiments. 
\cite{Johnson,Heinzel,Wang,YacobyPRL} 
It was revealed \cite{Johnson,Heinzel,Wang} that resonant
tunneling gives rise to a strong modulation of 
Coulomb blockade oscillations.
For a quantum dot embedded into a ring which was subjected to an 
Aharonov-Bohm flux,
the phase of the dot's transmission coefficient changed abruptly 
by $\pi$ across each resonance peak.\cite{YacobyPRL}
The last property was easily understood already in a 1D model
of non-interacting particles.\cite{YacobyPRB,Hackenbroich}
It is known that transport (conductance) and thermodynamic
(persistent current) characteristics of ballistic structures
have much in common. They both originate from coherent dynamics
of electrons in mesoscopic structures (quantum rings and dots).
However, persistent currents pertinent to a closed geometry, demonstrate
as a rule a more subtle behavior in comparison with conductance.
It is interesting to compare our predictions for 
the behavior of persistent currents
in double barrier ring with the results of
measurement of resonant electron tunneling in quantum dots.

First of all resonant tunneling gives rise to
a large (for a small barrier transparency) modulation of the persistent
current amplitude (see also Ref.~\onlinecite{Buttiker.Stafford}). 
This result demonstrates that resonant
tunneling influences drastically not only transport properties
of mesoscopic systems but the oscillating behavior of thermodynamic
quantities as well. In particular coherent resonant tunneling
in a closed system (a quantum ring) can be detected by measuring
the modulation of parity oscillations 
(see Fig.~\ref{fig:rectangular}).
Unlike resonant tunneling in an open geometry \cite{Johnson,Heinzel,Wang},
a quantum ring with two barriers exhibits a more complex behavior of resonance
peaks. For a compact ($l<<L$) double barrier structure, we have
true resonance oscillations (analogous to the ones observed
in Ref.~\onlinecite{Wang} for conductance). If the barriers are situated far 
apart from each other ($l \sim L/2$), the amplitude of the resonance peaks 
is determined by the single barrier current 
(semi-resonance behavior).\cite{Buttiker.Stafford}
The relative amplitude 
of true resonance peaks weakly depends on temperature (as for peaks
observed in Ref.~\onlinecite{Wang}), 
whereas  the semi-resonance peaks are easily
washed out by temperature.
The resonant behavior of persistent currents is sensitive to
correlation effects. It is known that electron-electron interactions
decrease the  effective barrier transparency (Kane-Fisher suppression).
This effect for weak interactions and low temperatures gives rise
to a more pronounced resonance structure. On the contrary, for
strong interactions or at high temperatures, the resonance behavior
of persistent current completely disappears.

We gratefully acknowledge discussions with C.~Canali, A.~Gogolin, L.~Gorelik, 
M.~Jonson, Yu~Lu, R.~Shekhter and V.~Shumeiko.
This work was supported by the Royal Swedish Academy of Science (KVA),
the Swedish Natural Science Research Council (NFR) and by INTAS
Grant No. 94-3962. 
One of us (I.K.) acknowledges the hospitality of the Department of 
Applied Physics, CTH/GU and of the International Centre for
Theoretical Physics, Trieste.

\end{document}